\documentclass[twocolumn,showpacs,preprintnumbers,amsmath,amssymb,superscriptaddress]{revtex4}

\usepackage{graphicx}
\usepackage{dcolumn}
\usepackage{bm}

\graphicspath{{./pic/}}

\begin{document}

\preprint{APS/123-QED}

\title{Electronic structure of 
Li$_{1+x}$[Mn$_{0.5}$Ni$_{0.5}$]$_{1-x}$O$_2$
studied by photoemission and x-ray absorption spectroscopy
}

\author{Y. Yokoyama}
\affiliation{Department of Physics, University of Tokyo, Chiba 277-8561, Japan}
\affiliation{Institute for Solid State Physics, University of Tokyo, Chiba 277-8581, Japan}
\author{D. Ootsuki}
\affiliation{Department of Physics, University of Tokyo, Chiba 277-8561, Japan}
\author{T. Sugimoto}
\affiliation{Department of Complexity Science and Engineering, University of Tokyo, Chiba 277-8561, Japan}
\author{H. Wadati}
\affiliation{Institute for Solid State Physics, University of Tokyo, Chiba 277-8581, Japan}
\author{J.~Okabayashi}
\affiliation{
Research Center for Spectrochemistry, University of Tokyo, Tokyo 113-0033, Japan 
}
\author{Xu Yang}
\affiliation{Key Laboratory of Physics and Technology for Advanced Batteries (Ministry of Education), College of physics, Jilin University, Changchun 130012, People's Republic of China}
\author{Fei Du}
\affiliation{Key Laboratory of Physics and Technology for Advanced Batteries (Ministry of Education), College of physics, Jilin University, Changchun 130012, People's Republic of China}
\author{Gang Chen}
\affiliation{Key Laboratory of Physics and Technology for Advanced Batteries (Ministry of Education), College of physics, Jilin University, Changchun 130012, People's Republic of China}
\author{T. Mizokawa}
\affiliation{Department of Applied Physics, Waseda University, Tokyo 169-8555, Japan}

\date{\today}

\begin{abstract}
We have studied the electronic structure of
Li$_{1+x}$[Mn$_{0.5}$Ni$_{0.5}$]$_{1-x}$O$_2$ ($x$ = 0.00 and 0.05),
one of the promising cathode materials for Li ion battery,
by means of x-ray photoemission and absorption spectroscopy.
The results show that the valences of Mn and Ni are basically 4+ and 2+, respectively.
However, the Mn$^{3+}$ component in the $x$ = 0.00 sample gradually increases 
with the bulk sensitivity of the experiment, indicating that the Jahn-Teller active
Mn$^{3+}$ ions are introduced in the bulk due to the site exchange between Li and Ni.
The Mn$^{3+}$ component gets negligibly small in the $x$ = 0.05 sample,
which indicates
that the excess Li suppresses the site exchange and removes the Jahn-Teller active
Mn$^{3+}$.
\end{abstract}

\maketitle

\newpage

Layered transition-metal oxides, in which transition-metal $M$
atoms form two-dimensional triangular lattices with 
edge-sharing $M$O$_6$ octahedra, are known as cathode materials 
for rechargeable batteries.
In particular, Li$_x$CoO$_2$ has been widely studied 
for the practical use as a positive electrode material 
in commercial Li ion batteries \cite{Mizushima}.
Rechargeable batteries using Li$_x$CoO$_2$ cathodes
exhibit the highest performance among batteries 
using similar transition metal oxides \cite{Mizushima,Plichta,Gibbard,Nagura}.
However, Co is relatively expensive and alternative cathode materials 
should be developed with less expensive transition-metal elements. 
Since Mn and Ni are less expensive than Co, LiMn$^{4+}_{0.5}$Ni$^{2+}_{0.5}$O$_2$ 
is one of the promising materials \cite{Ohzuku, Cushing}.
LiMn$_{0.5}$Ni$_{0.5}$O$_2$ has the layered $\alpha$-NaFeO$_2$ structure 
(rhombohedral system, space group  {\it R$\bar{3}$m}), and consists
of the $M$O$_2$ ($M$ = Mn and Ni) layers and the interlayers of Li ions.
The Li ions occupy the octahedral sites between the $M$O$_2$ layers.
In the charging process, Ni$^{2+}$ ejects two electrons and changes to Ni$^{4+}$.
By increasing the ratio of Li in Li$_{1+x}$[Mn$_{0.5}$Ni$_{0.5}$]$_{1-x}$O$_2$, 
the capacity retention is improved as $x$ increases up to 0.05 \cite{Idemoto}.
If Jahn-Teller active Mn$^{3+}$ species exist, the capacity retention 
would be degraded due to the Jahn-Teller distortion of Mn$^{3+}$O$_6$ octahedron.
Moreover, the exchange between the Ni$^{2+}$ and Li$^+$ ions plays important roles \cite{Idemoto}.
In order to clarify the effect of the excess Li ions on the electronic structure,
we have performed x-ray photoemission spectroscopy (XPS) and x-ray absorption spectroscopy (XAS) 
measurements of Li$_{1+x}$[Mn$_{0.5}$Ni$_{0.5}$]$_{1-x}$O$_2$.


Powder samples of Li$_{1+x}$[Mn$_{0.5}$Ni$_{0.5}$]$_{1-x}$O$_2$ 
($x$ = 0.00 and 0.05) were grown by solid state reaction.
The samples were pressed onto the carbon tapes which were
attached to the sample holder, and were introduced to the chambers
under the ultrahigh vacuum.
XPS measurements were performed by using JEOL JPS-9200
with a Mg K$\alpha$ x-ray source (1253.6 eV).
XAS measurements were performed at beamline 7A, Photon Factory, KEK. 
In the total electron yield (TEY) mode, the total electrons were counted
by measuring the sample current. In the fluorescence yield (FY) mode, 
the O 1$s$ x-ray emission was measured for the inverse partial fluorescence yield (IPFY) method \cite{Wadati_2010,Wadati_2012}.


Figures 1(a) and (b) show the Ni 2$p$ and Mn 2$p$ XPS spectra of
Li$_{1+x}$[Mn$_{0.5}$Ni$_{0.5}$]$_{1-x}$O$_2$ ($x$ = 0.00 and 0.05).
In the Ni 2$p$ XPS,
the peak around 860 eV corresponds to Ni 2$p_{3/2}$
and the peak around 880 eV corresponds to Ni 2$p_{1/2}$.
In the Mn 2$p$ XPS, the peaks around 643 eV and 655 eV correspond
to the Mn 2$p_{3/2}$ and Mn 2$p_{1/2}$ branches, respectively.
The peak shift with $x$ is not observed in the Mn 2$p$ and Ni 2$p$
core levels, showing that the valences of Mn and Ni do not
depend on $x$ appreciably.

In Fig. 2, the Ni 2$p$ XPS is compared with the configuration interaction
(CI) calculation on a NiO$_6$ clsuter model with $\Delta$ = 4.5 eV, 
$(pd\sigma)$ = $-1.5$ eV, and  $U$ = 7.0 eV.
In the present cluster model, the ground state with $^3A_{2g}$ symmetry is 
given by the linear combination of $d^8$, $d^9L$, and $d^{10}L^2$ configurations,
where $L$ denotes an O 2$p$ hole. The energy difference between $d^8$ and $d^9L$
corresponds to $\Delta$ and that between $d^9L$ and $d^{10}L^2$ is basically given
by $\Delta+U$. The final states are described by the linear combinations of 
$cd^8$, $cd^9L$, and $cd^{10}L^2$ configurations, where $c$ denotes an Ni 2$p$ hole.
The Coulomb interaction between the Ni 3$d$ electrons are given by the Slater
integrals $F^0(3d,3d)$, $F^2(3d,3d)$, and $F^4(3d,3d)$.
The average Ni 3$d$-Ni 3$d$ Coulomb interaction $U$
is expressed by $F^0(3d,3d)$ and is an adjustable parameter.
$F^2(3d,3d)$ and $F^4(3d,3d)$ are fixed to 80\% of
the atomic Hartree-Fock values \cite{deGroot}.
The Coulomb interaction between the Ni 2$p$ core hole and the Ni 3d electron
is expressed by the Slater integrals $F^0$(2p,3d),
$F^2(2p,3d)$, and $G^1(2p,3d)$.
The average Ni 2$p$-Ni 3$d$ Coulomb interaction $Q$
is expressed by $F^0(2p,3d)$ and is fixed to $U/0.8$.
$F^2(2p,3d)$ and $G^1(2p,3d)$ are fixed to 80\% of
the atomic Hartree-Fock values \cite{deGroot}.
The energy difference between $cd^8$ and $cd^9L$
corresponds to $\Delta-Q$ and that between $cd^9L$ and $cd^{10}L^2$
is $\Delta-Q+U$.
The transfer integrals between the Ni 3$d$ and O 2$p$ orbitals
are given by $(pd\sigma)$ and $(pd\pi)$ where the ratio
$(pd\sigma)$/$(pd\pi)$ is fixed at -2.16.
As shown in Fig. 2, the charge-transfer satellite is well reproduced
by the calculation. The obtained parameters satisfy $\Delta < U$, 
indicating that the Ni$^{2+}$ state falls in the charge-transfer regime. 
Since $\Delta-Q$ is negative, the main and satellite peaks
are dominated by the $cd^9L$ and $cd^8$ configurations, respectively.

\begin{figure}
  \begin{center}
    \includegraphics[width=0.45\textwidth]{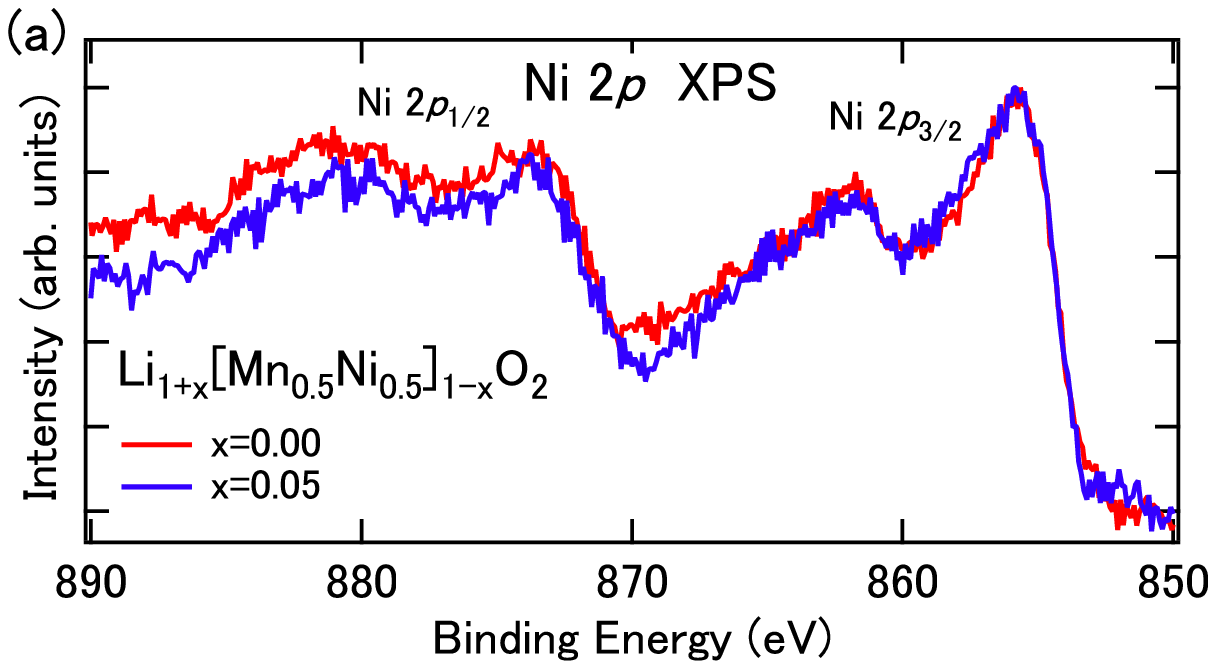}
    \includegraphics[width=0.45\textwidth]{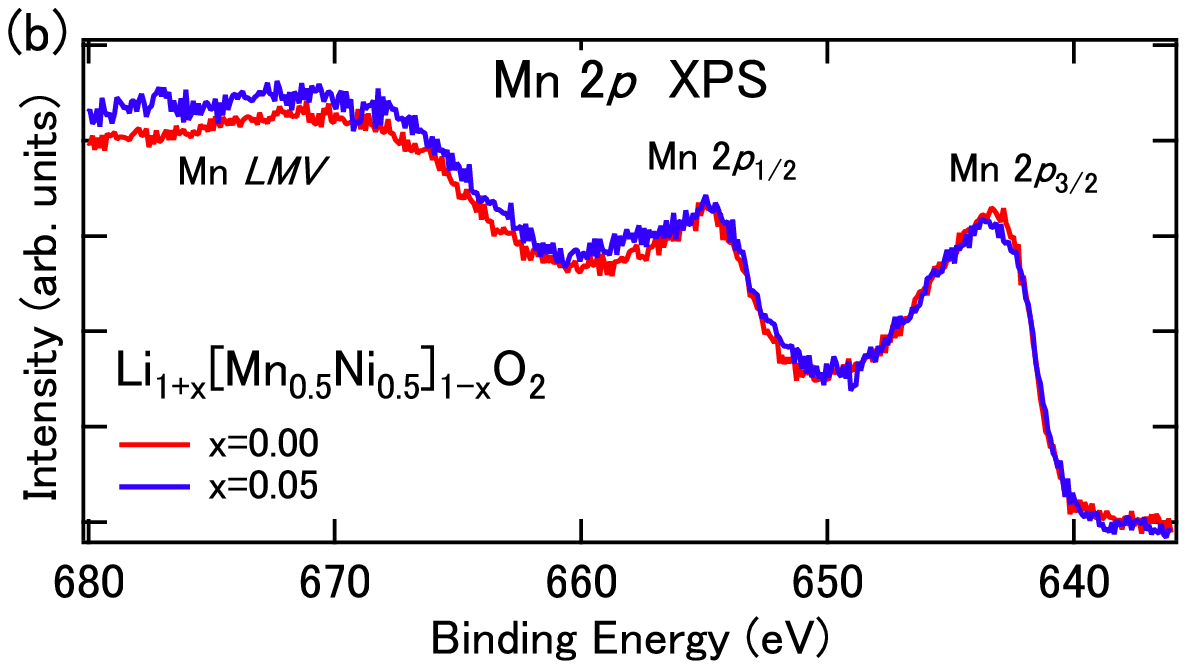}
  \end{center}
  \caption{
(a) Ni 2$p$ and (b) Mn 2$p$ XPS of
Li$_{1+x}$[Mn$_{0.5}$Ni$_{0.5}$]$_{1-x}$O$_2$ ($x$ = 0.00 and 0.05).
}
\end{figure}

\begin{figure}
  \begin{center}
    \includegraphics[width=0.45\textwidth]{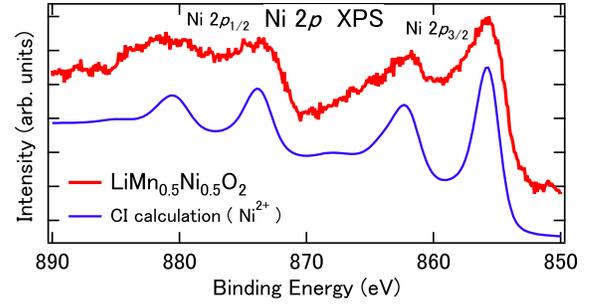}
  \end{center}
  \caption{
Calculated Ni 2$p$ XPS spectrum (blue line) compared 
with the Ni 2$p$ XPS of LiMn$_{0.5}$Ni$_{0.5}$O$_2$ (red line).}
\end{figure}

\begin{figure}
\begin{center}
   \includegraphics[width=0.45\textwidth]{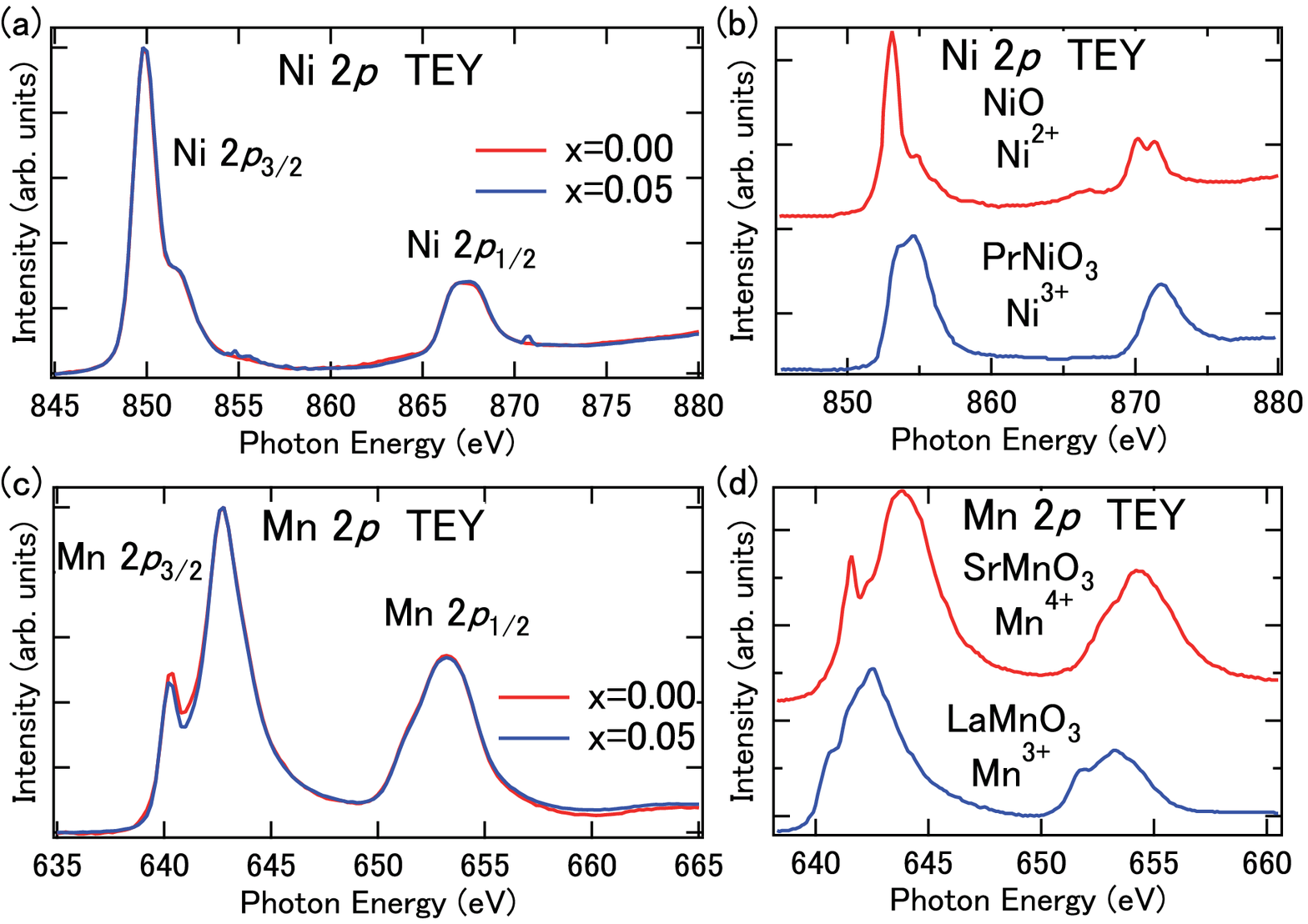}
   \includegraphics[width=0.45\textwidth]{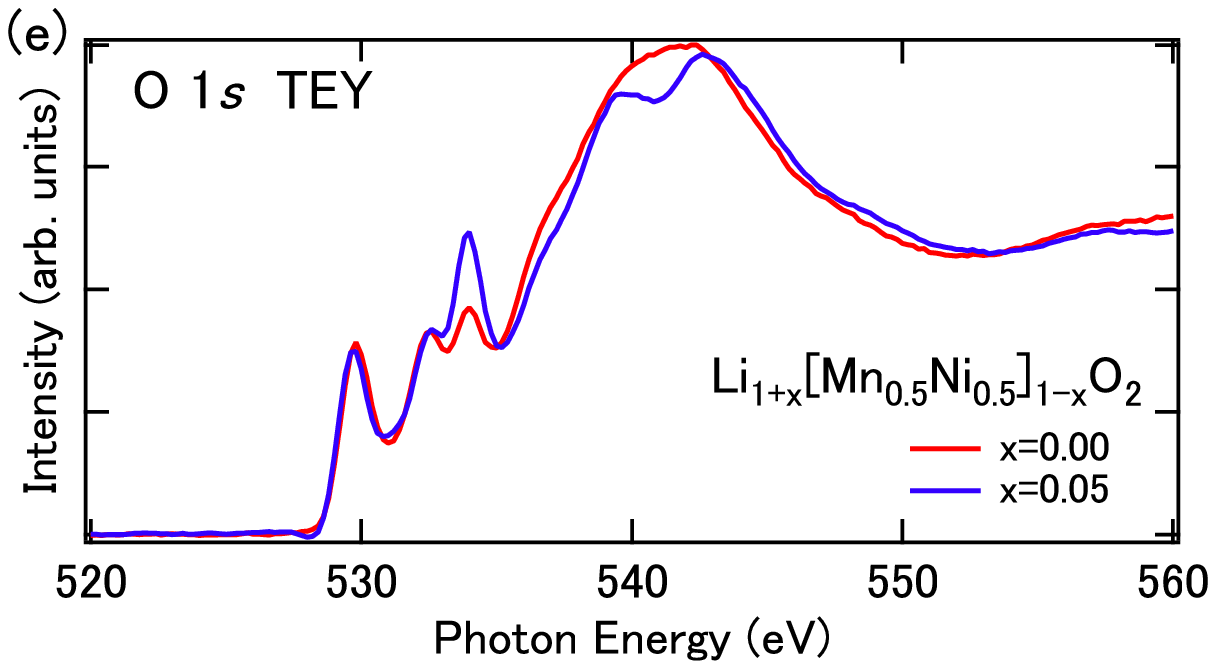}
\end{center}
  \caption{
(a) Ni 2$p$ XAS spectra of
Li$_{1+x}$[Mn$_{0.5}$Ni$_{0.5}$]$_{1-x}$O$_2$ ($x$ = 0.00 and 0.05) and (b) their comparison with the reference data \cite{Medarde}. (c) Mn 2$p$ XAS spectra of
Li$_{1+x}$[Mn$_{0.5}$Ni$_{0.5}$]$_{1-x}$O$_2$ and (d) their comparison with the reference data \cite{Vasiliev}.
(e) O 1$s$ XAS spectra of Li$_{1+x}$[Mn$_{0.5}$Ni$_{0.5}$]$_{1-x}$O$_2$.
}
\end{figure}

We performed XAS measurements with various $x$ in the TEY mode.
The probing depth of the TEY XAS measurement is larger than that of the XPS measurement.
Figure 3(a) and (b) show the Ni 2$p$ TEY XAS spectra and their comparison with 
the reference data \cite{Medarde}. The peak around 850 eV corresponds to the Ni 2$p_{3/2}$ 
edge and the peak around 867 eV corresponds to the Ni 2$p_{1/2}$ edge. 
The lineshape shows that the valence of Ni is dominated by 2+ and 
does not change as the ratio of Li increases.

In the Mn 2$p$ XAS shown in Fig. 3(c), the peak around 640 eV corresponds 
to the transition from the Mn 2$p_{3/2}$ core level to the Mn 3$d$ $e_g$ state 
and the peak around 643 eV corresponds to the transition from the Mn 2$p_{3/2}$ 
core level to the Mn 3$d$ $t_{2g}$ state. The other peak around 653 eV 
corresponds to the Mn 2$p_{1/2}$ edge. The lineshape indicates that 
the valence of Mn is almost 4+. However, it is observed that the height of 
the $e_g$ edge relative to the $t_{2g}$ edge gets slightly lower as the ratio of Li increases.
Considering the fact that the Ni 2$p$ spectral change is negligible,
the Mn 2$p$ spectral change with $x$ should be taken into account seriously.

Figure 3(e) shows the O 1$s$ TEY XAS spectra that represent the transitions 
from the O 1$s$ core level to the O 2$p$ orbitals mixed into the unoccupied Mn and Ni 3$d$ states. 
Interestingly, the peak around 533 eV gains its intensity with $x$ and would be related 
to the Mn valence change observed in the Mn 2$p$ XAS.

Figure 4(a) shows the Ni 2$p$ TEY XAS spectrum and its comparison with the CI calculation. 
By using the parameters obtained in Fig. 2, we calculated the theoretical spectrum and 
attempted to reproduce the experimental result. The theoretical spectrum (calculated by 
$\Delta$ = 4.5 eV, $pd\sigma$ = $-1.5$ eV, and $U$ = 7.0 eV) is in good agreement with 
the experimental spectrum, indicating that the Ni 2$p$ TEY XAS spectra are consistent 
with the Ni 2$p$ XPS spectra.
In the Ni 2$p$ XAS, the final states are described by the linear combinations of 
$cd^9$ and $cd^{10}L$ configurations.
Since $\Delta-Q+U$ is positive, the main peak is dominated
by the $cd^9$ configurations and the satellite peak is very small.
In this sense, the charge transferred configurations are irrelevant
in the case of Ni 2$p$ XAS.

Figure 4(b) shows the Mn 2$p$ TEY XAS spectrum and its comparison with the calculated spectra 
for the transitions from $d^3$ to $cd^4$. Since $\Delta-Q+U$ is positive and the charge transfer 
satellite is not observed, the Mn 2$p$ XAS lineshape can be evaluated without including 
the charge transferred configurations. Instead of the O 2$p$-Mn 3$d$ transfer integrals, 
the energy splitting between the $e_g$ and $t_{2g}$ orbitals (10$Dq$) is adjusted to 
reproduce the experimental result. The estimated value of 10$Dq$ is 2.0 or 2.5 eV as shown in Fig. 4(b).
By using 10$Dq$ = 2.5 eV, we calculated the Mn$^{3+}$ spectrum for the transitions from $d^4$ to $cd^5$. 
Then, we added the spectrum of Mn$^{3+}$ to that of Mn$^{4+}$, trying to reproduce the experimental 
spectra of Li$_{1+x}$[Mn$_{0.5}$Ni$_{0.5}$]$_{1-x}$O$_2$ ($x$ = 0.00 and 0.05).
Figure 4(c) shows the Mn 2$p$ XAS spectra and their comparison with the
theoretical spectra which correspond to the mixed valence of Mn$^{4+}$ and Mn$^{3+}$.
In the experimental result, the intensity of $e_g$ peak ($\sim$ 640.2 eV) 
relative to that of $t_{2g}$ peak ($\sim$ 642.7 eV) is reduced with $x$.
The theoretical spectra indicate that the intensity of $e_g$ peak ($\sim$ 640.5 eV) 
relative to that of $t_{2g}$ peak ($\sim$ 642.5 eV) gets stronger as the ratio of Mn$^{3+}$ 
increases. Therefore, it is considered that the Mn$^{3+}$ ions exist at $x$ = 0.00 
and the amount of Mn$^{3+}$ decreases with $x$.

\begin{figure}
\includegraphics[width=0.45\textwidth]{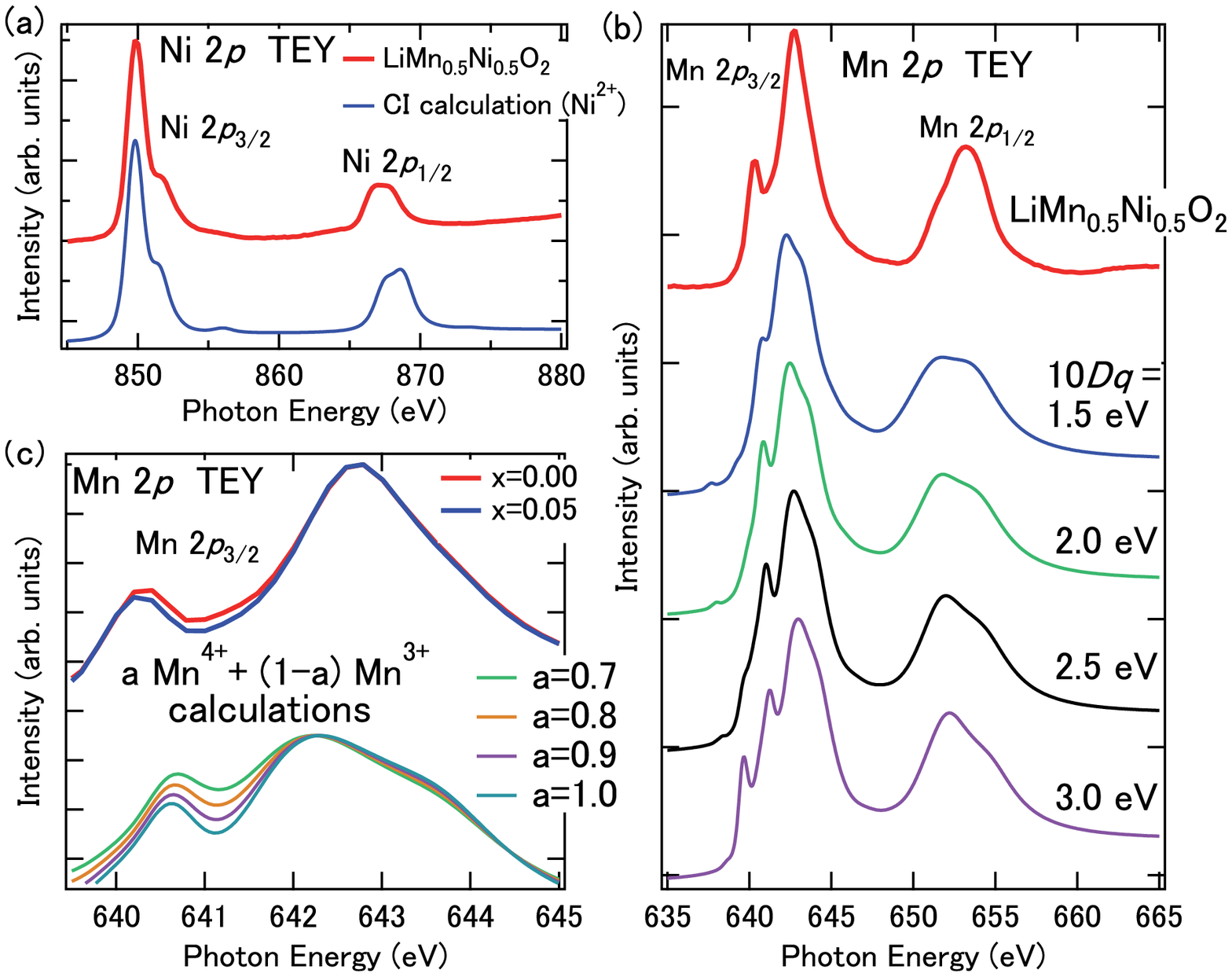}
  \caption{
(a) Calculated Ni 2$p$ XAS spectrum (blue line) compared 
with the Ni 2$p$ XPS of LiMn$_{0.5}$Ni$_{0.5}$O$_2$ (red line).
(b) Calculated Mn 2$p$ XAS spectra as a function
of $10Dq$. The calculated results are compared
with the Mn 2$p$ XPS of LiMn$_{0.5}$Ni$_{0.5}$O$_2$ (red line).
(c)  Mn 2$p$ XAS spectra of Li$_{1+x}$[Mn$_{0.5}$Ni$_{0.5}$]$_{1-x}$O$_2$ ($x$ = 0.00 and 0.05) and their comparison with the theoretical spectra obtained by adding the calculated spectrum of Mn$^{4+}$ and that of Mn$^{3+}$.
}
\end{figure}

\begin{figure}
  \begin{center}
    \includegraphics[width=0.45\textwidth]{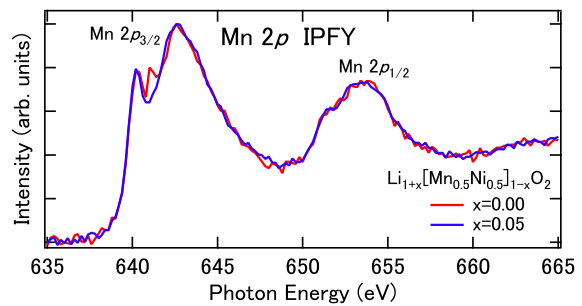}
  \end{center}
  \caption{
IPFY Mn 2$p$ XAS of Li$_{1+x}$[Mn$_{0.5}$Ni$_{0.5}$]$_{1-x}$O$_2$ ($x$ = 0.00 and 0.05)}
\end{figure}

Figure 5 shows the IPFY XAS spectra of Mn 2$p$.
The peak around 640 eV corresponds to the transition from the Mn 2$p_{3/2}$ core level
to the Mn 3$d$ $e_g$ state and the peak around 643 eV corresponds to the transition 
from the Mn 2$p_{3/2}$ core level to the Mn 3$d$ $t_{2g}$ state. The other peak 
around 653 eV corresponds to the Mn 2$p_{1/2}$ edge. The sharp peak around 640 eV
is specific for Mn$^{4+}$ where the $e_g$ orbitals are unoccupied. In addition, 
we observed a small peak between the peaks around 640 eV and 643 eV at $x$ = 0.00. 
The small peak corresponds to the Mn 2$p_{3/2}$ edge of Mn$^{3+}$, indicating that
the $x$ = 0.00 sample contains the Mn$^{3+}$ ions at the bulk. The small peak 
vanished with increasing $x$, suggesting that the Mn$^{3+}$ ions change into 
the Mn$^{4+}$ ions as the ratio of Li increases up to x = 0.05.

The Ni 2$p$ and Mn 2$p$ XPS spectra, which are very surface sensitive,
indicate that the Ni and Mn valence states do not change at the surface
between $x$ = 0.00 and 0.05
in Li$_{1+x}$[Mn$_{0.5}$Ni$_{0.5}$]$_{1-x}$O$_2$.
The small contribution of Mn$^{3+}$ is suggested for $x$ = 0.00
in the Mn 2$p$ TEY XAS
which is more bulk sensitive than the Mn 2$p$ XPS.
The most bulk sensitive IPFY XAS spectra indicate that the Jahn-Teller
active Mn$^{3+}$ component exists at $x$ = 0.00 and is dramatically suppressed
by increasing x.
The existence of Mn$^{3+}$ at $x$ = 0.00 can be attributed to the site exchange
between Ni and Li. The fact that the Mn$^{3+}$ peak is removed with $x$ indicates
that the almost all the Mn$^{3+}$ ions are changed to Mn$^{4+}$ by introducing
the excess Li up to x = 0.05 and probably by suppressing the site exchange.
It is considered that the Jahn-Teller active Mn$^{3+}$ ions
at the bulk deteriorate the cycle performance of the Li ion battery. 
Therefore, the decrease of Mn$^{3+}$ or the decrease of the site exchange 
with $x$ is useful to improve the performance.

The present work has been performed under the approvals of the Photon Factory Program Advisory Committee (Proposal No. 2013G680).

\end{document}